\documentstyle[11pt,epsf]{article}
\def\beq{\begin{equation}}
\def\eeq{\end{equation}}
\def\beqa{\begin{eqnarray}}
\def\eeqa{\end{eqnarray}}

\def\ie{{\it i.e.,\ }}

\def\slfrac#1/#2{\leavevmode\kern.1em\raise.5ex\hbox{\the\scriptfont0
         #1}\kern-.1em/\kern-.15em\lower.25ex\hbox{\the\scriptfont0 #2}}
\def\gev{{\rm\,GeV}}
\def\tev{{\rm\,TeV}}

\def\bbar{{\bar b}}
\def\pbar{{\bar p}}
\def\qbar{{\bar q}}

\def\tbar{{\bar t}}

\def\khat{{\hat k}}
\begin{document}

\begin{titlepage}
\samepage{
\setcounter{page}{1}
\rightline{UFIFT--HEP-97-31}
\vskip1in
\begin{center}
{\Large\bf Using Collider Event Topology in the Search for the \\  Six-Jet Decay of Top Quark-Antiquark Pairs\footnote{Supported in part by U.S.~Department of Energy grant DE--FG05--86ER--40272.}}
\vskip0.5in
{\large R. D. Field and Y. A. Kanev \\  {\em Institute for Fundamental Theory, Department of Physics, \\
University of Florida, Gainesville, FL 32611}}
\vskip0.5in
{\large November 1997}
\end{center}
\vskip1in
\begin{abstract}
We investigate the use of the event topology as a tool in the search for the six-jet decay of top-pair production in proton-antiproton collisions at $1.8\tev$. Modified Fox-Wolfram ``shape" variables, ${\hat H}_\ell$, are employed to help distinguish the top-pair signal from the ordinary QCD multi-jet background.  The ${\hat H}_\ell$'s can be constructed directly from the calorimeter cells or from jets. Events are required to lie in a region of ${\hat H}_\ell$-space defined by $L_\ell < {\hat H}_\ell < R_\ell$ for $\ell=1,\ldots,6$, where the left, $L_\ell$, and right, $R_\ell$, cuts are determined by a genetic algorithm (GA) procedure to maximize the signal over the square root of the background.  We are able to reduce the background over the signal to less than a factor of $100$ using purely topological methods without using jet multiplicity cuts and without the aid of  b-quark tagging.   
\end{abstract}
\smallskip
}
\end{titlepage}

\setcounter{footnote}{0}

\section{Introduction}

The challenge at hadron colliders is to disentangle any new physics that may be present from the ``ordinary'' QCD background.  Hadron collider events can be very complicated and quite often one has the situation where the signal is hiding beneath the background.  In addition, there are many variables that describe a high energy collider event and it is not always obvious which variables best isolate the signal or precisely what data selection (or cuts) optimally enhance the signal over the background.  In this paper, we use information on the event topology to help enhance the signal over the background.  We define six modified Fox-Wolfram ``shape" variables, ${\hat H}_\ell$, $\ell=1,\ldots,6$, to characterize the topology of the event.  The ${\hat H}_\ell$'s can be constructed directly from the calorimeter cells or from the jets.  To illustrate our techniques, we will attempt to isolate the six-jet decay of top-pair production in proton-antiproton collisions at $1.8\tev$ from the ordinary QCD multi-jet background, without tagging b-quarks.  B-quark tagging would, of course, further enhance the signal to background ratio. 

\begin{figure}[htbp]
  \begin{center}
    \leavevmode
    \epsfxsize=4in 
    \epsfbox{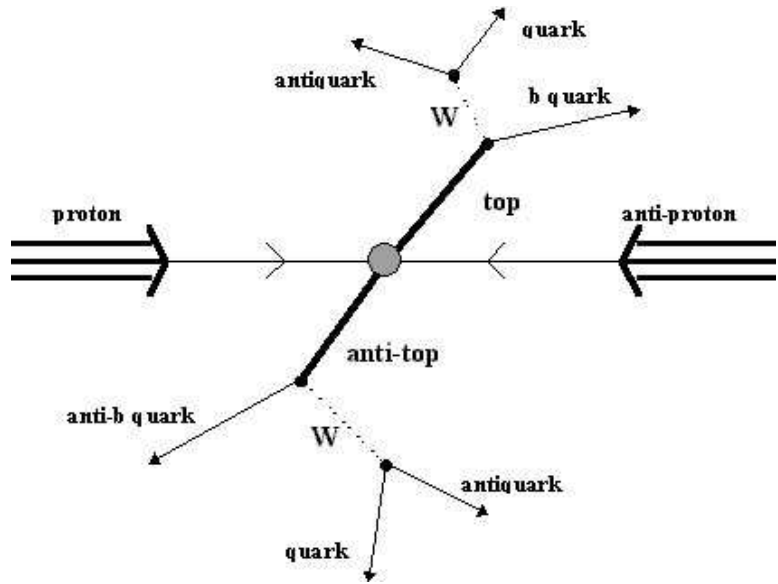}
  \end{center}
\caption{\footnotesize Illustration of top-pair production in proton-antiproton collisions in which both of the $W$ bosons decay hadronically resulting in a final state consisting of a $b\bbar$ pair and two $q\qbar$ pairs.}
\label{Fig:sig1}
\end{figure}

\begin{figure}[htbp]
  \begin{center}
    \leavevmode
    \epsfxsize=4in 
    \epsfbox{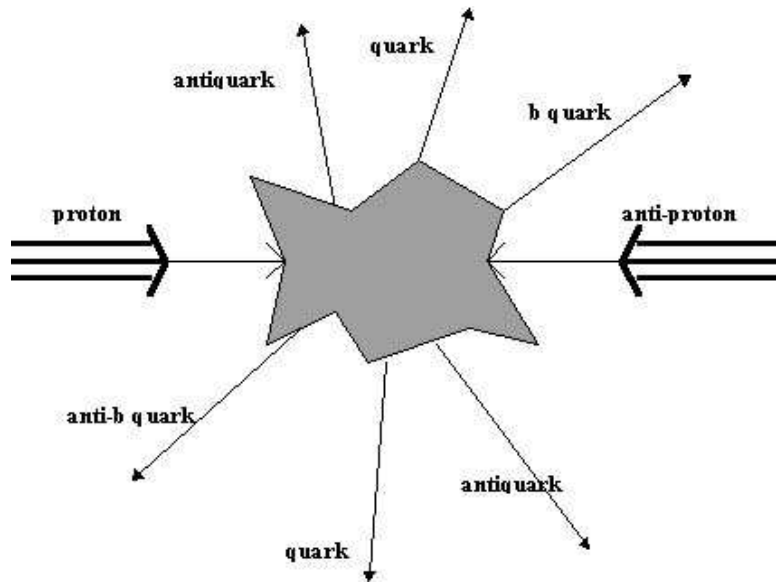}
  \end{center}
\caption{\footnotesize Shows the event topology for the top-pair signal.  If each of the outgoing partons produces a distinct jet, then the final state contains six jets.}
\label{Fig:sig2}
\end{figure}

\begin{figure}[htbp]
  \begin{center}
    \leavevmode
    \epsfxsize=4in 
    \epsfbox{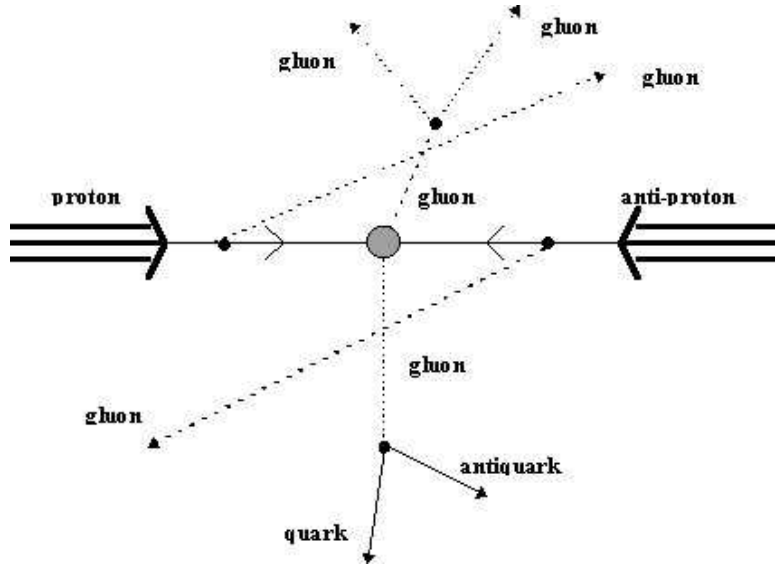}
  \end{center}
\caption{\footnotesize Illustration of the QCD multi-jet background to the top-pair production in proton-antiproton collisions shown in Fig.~\ref{Fig:sig1}.}
\label{Fig:bak}
\end{figure}

The top quark decays into a $b$-quark and a $W$ boson ($t\to bW$).  The $W$ boson decays into a lepton ($e$ or $\mu$) and a neutrino about $22\%$ ($\slfrac2/9$) of the time and into a quark-antiquark pair about $67\%$ ($\slfrac6/9$) of the time.  This implies that when top-pairs are produced in 
hadron-hadron collisions, $p\pbar\to t\tbar$+X, both of the $W$ bosons decay into a lepton and neutrino only about $5\%$ of the time resulting in the final state consisting of two leptons, two neutrinos, and two b-quarks ($\ell\ell\nu\nu b\bbar$).  This distinctive final state constitutes the ``discovery" mode of the top quark at hadron colliders \cite{CDF_1,D0_1}.  On the other hand, it is considerable more likely for one of the $W$ bosons to decay into a quark-antiquark pair resulting in a final state consisting of  a lepton, a neutrino, a $b\bbar$, and a $q\qbar$ pair.  The $\ell\nu b\bbar q\qbar$ mode occurs about $35\%$ of the time or about $7$ times more often than the purely leptonic mode.  The backgrounds are larger for this decay mode, but so is the signal.  When each of the four outgoing quarks produce a distinct jet, the resulting event contains a lepton, a neutrino, and four jets ($\ell\nu jjjj$). This decay mode is used to analyze the properties of the top quark in more detail and to determine, for example, the top mass \cite{CDF_2, D0_2, D0_3}.  The purely hadronic decay mode shown in Fig.~\ref{Fig:sig1} occurs about $60\%$ of the time, and produces the ``six-jet" topology shown in Fig.~\ref{Fig:sig2}.  The six-jet decay mode of top-pair production is buried underneath ``ordinary" QCD multi-jet production such as that illustrated in Fig.~\ref{Fig:bak}.

We will attempt to isolate the $t\tbar$ six-jet mode from the background using only the event topology.   The signal in Fig.~\ref{Fig:sig1} contains $b$ quarks whereas the QCD multi-jet background in Fig.~\ref{Fig:bak}, in general, does not.  Therefore,  $b$-quark tagging will improve the signal to background ratio. However, we would like to investigate how well one can do using only the event topology. We begin our analysis of the signal and background in Section II with a discussion of the event simulation and detection.   In Section III, we define the ${\hat H}_\ell$ variables that characterize the collider event topology and in Section IV we discuss the genetic algorithm (GA) that we use to find optimal regions of ${\hat H}_\ell$-space.  We reconstruct the top-pair invariant mass in Section V and in Section VI we isolate the top-pair topology by making ${\hat H}_\ell$ cuts. Section VII is reserved for summary and conclusions.

\vskip 0.2in
\section{Event Simulation and Detection}

ISAJET version 7.06 \cite{isajet} is used to generate  top quarks 
with a mass of $175\gev$ in $1.8\tev$ proton-antiproton collisions.  At this energy, $175\gev$ top-pairs are produced via quark-antiquark annihilation, $q\qbar\to t\tbar$,  about $88\%$ of the time and by gluon-gluon fusion, $gg\to t\tbar$, the remaining $12\%$. We refer to this as the ``signal". We have normalized the top cross section to be $7.5$ pb corresponding to $750$ events with an integrated luminosity of $100$/pb \cite{CDF_2,D0_2,D0_3}.  The ``background" consists of ordinary QCD multi-jet events generated using ISAJET with the hard-scattering transverse momentum, $\khat_T$, greater than $20\gev$.  ISAJET uses the ``leading pole" approximation to produce multi-jets and not the exact matrix elements.  (The $2\to2$ matrix elements are exact but not the $2\to N$ with $N>2$.)  Because of this the precise numbers in this paper should not be taken too seriously.  Nevertheless, ISAJET is sufficient to illustrate our techniques. 

We do not attempt to do a detailed simulation of  the CDF or D0 detector \cite{CDF_2, D0_2}.  Events are analyzed by dividing the solid angle into ``calorimeter'' cells having size $\Delta\eta\Delta\phi=0.1\times7.5^\circ$, where $\eta$ and $\phi$ are the pseudorapidity and azimuthal angle, respectively.  Our simple calorimeter covers the range $|\eta|<4$ and has $3840$ cells.  A single cell has an energy (the sum of the energies of all the particles that hit the cell {\em excluding} neutrinos) and a direction given by the coordinates of the center of the cell.  The transverse energy of each cell is computed from the cell energy and direction. We have taken the energy resolution to be perfect, which means that the only resolution effects are caused by the lack of spatial resolution due to the cell size.

\vskip 0.2in
\section{Variables that Characterize the Event Topology}

\subsection{Fox-Wolfram Moments}
In $1979$ Geoffrey Fox and Stephen Wolfram \cite{Fox} constructed a complete set of rotationally invariant observables, $H_\ell$, which can be used to characterize the ``shapes" of the final states in electron-positron annihilations.  They are constructed from the momentum vectors, $\vec p$, of  all the final state particles as follows,
\beq
H_\ell=\left({4\pi\over 2\ell+1}\right)\sum^{+\ell}_{m=-\ell}
\Bigl|\sum^{particles}_i Y^m_\ell(\Omega_i){|{\vec p}_i|\over E_{tot}}\Bigr|^2,\eeq
where the inner sum is over the particles produced and $Y^m_\ell$ are the spherical harmonics.  Here one must choose a particular set of axes to evaluate the angles,
$\Omega_i=(\theta_i,\phi_i)$, of the final state particles, but the values of the $H_\ell$ are independent of this choice.  These moments lie in the range $0\le H_\ell \le 1$ and
if energy conserved in the final state then $H_0=1$ ({\em neglecting the masses}).  If momentum is conserved in the final state then $H_1=0$.

The Fox-Wolfram observables (or moments) constitute a complete set of shape parameters.  For example, the
collinear ``two-jet" final state results in $H_\ell\approx 1$ for even $\ell$ and $H_\ell\approx 0$ for odd $\ell$.  Events that are completely spherically symmetric give $H_\ell\approx 0$ for all $\ell$.  

\subsection{Constructing Fox-Wolfram Moments from Calorimeter Cells}
In hadron-hadron collisions spherical symmetry is lost and we are interested more in the shape of events in the transverse plane.  For example, the Fox-Wolfram moments when applied directly to hadron-hadron collisions would interpret a minimum bias event as a ``two-jet" event, whereas we would like to have a minimum bias event treated more like a spherically symmetric $e^+e^-$ final state (\ie no structure).  To accomplish this, we define the following modified
Fox-Wolfram moments for hadron-hadron collisions,
\beq
{\hat H}_\ell(cell)=\left({4\pi\over 2\ell+1}\right)
\sum^{+\ell}_{m=-\ell}
\Bigl|\sum_i^{cells} Y^m_\ell(\Omega_i){E_T^i\over E_T({\rm sum})}\Bigr|^2,\eeq
where the inner sum is over all the calorimeter cells in the event with transverse energy, $E_T^i$, greater than some minimum (for example,  $5\gev$) and $\Omega_i=(\theta_i,\phi_i)$ are the angular locations of the center of the cell. In this case, $E_T(sum)$ is the total transverse energy of all the cells that are included in the sum. The calorimeter cells contain all the information concerning the topology of the event and it is not necessary to define jets.  These modified moments also lie in the range $0\le {\hat H}_\ell \le 1$ and by definition ${\hat H}_0=1$.

\begin{table}[htbp]
\begin{center}
\begin{tabular}{||l||c|c||} \hline \hline
 & ${\hat H}_\ell(cell)$ & ${\hat H}_\ell(jet)\ R_j\!=\!0.4$  \\ 
 & \ Signal \ \ \ Background & \ Signal \ \ \ Background  \\ \hline\hline
 ${\hat H}_1$ & {\small 0.2053$\pm$0.0797 \ \ 0.3160$\pm$0.1698} 
&{\small  0.1970$\pm$0.0784 \ \ 0.3104$\pm$0.1748}   \\ \hline
 ${\hat H}_2$ & {\small 0.2827$\pm$0.1093 \ \ 0.5479$\pm$0.3581} 
& {\small 0.2711$\pm$0.1046 \ \ 0.5557$\pm$0.3737}   \\ \hline
 ${\hat H}_3$ & {\small 0.2670$\pm$0.0951 \ \ 0.3849$\pm$0.1883} 
& {\small 0.2593$\pm$0.0934 \ \ 0.3890$\pm$0.1985}   \\ \hline
 ${\hat H}_4$ & {\small 0.2738$\pm$0.0959 \ \ 0.4774$\pm$0.2670} 
& {\small 0.2713$\pm$0.0976 \ \ 0.4937$\pm$0.2894}   \\ \hline
 ${\hat H}_5$ & {\small 0.2688$\pm$0.0908 \ \ 0.4058$\pm$0.1946} 
& {\small 0.2723$\pm$0.0964 \ \ 0.4223$\pm$0.2150}   \\ \hline
 ${\hat H}_6$ & {\small 0.2640$\pm$0.0867 \ \ 0.4463$\pm$0.2296} 
& {\small 0.2744$\pm$0.0965 \ \ 0.4738$\pm$0.2612}   \\ \hline\hline
\end{tabular}
\end{center}
\caption{\footnotesize Shows the mean value and standard deviation from the mean (mean$\pm\sigma$) of six of the Modified Fox-Wolfram moments, ${\hat H}_\ell$, constructed from the calorimeter cells (with $E_T(cell)\!>\!5\gev$) and from jets with $R_j\!=\!0.4$ and $E_T(jet)\!>\!15\gev$.  Results are shown for the top-pair signal and the QCD multi-jet background in $1.8\tev$ proton-antiproton collisions.}
\label{Table:mean}
\end{table}

\begin{figure}[htbp]
  \begin{center}
    \leavevmode
    \epsfxsize=4in 
    \epsfbox{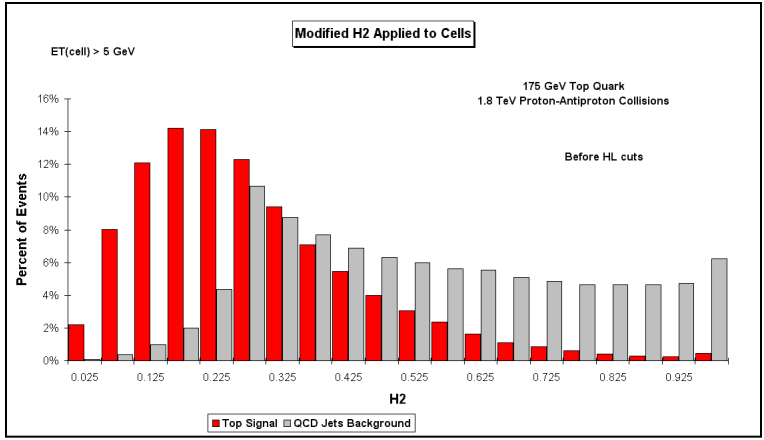}
  \end{center}
\caption{\footnotesize Shows the modified Fox-Wolfram moment, ${\hat H}_2$, calculated directly from the calorimeter cells with $E_T(cell)\!>\!5\gev$ for top-pair signal and for the QCD multi-jet background.  The plot shows the percentage of events in a $0.05$ bin with the sum of all bins normalized to $100\%$.}
\label{Fig:H2}
\end{figure}

\begin{figure}[htbp]
  \begin{center}
    \leavevmode
    \epsfxsize=4in 
    \epsfbox{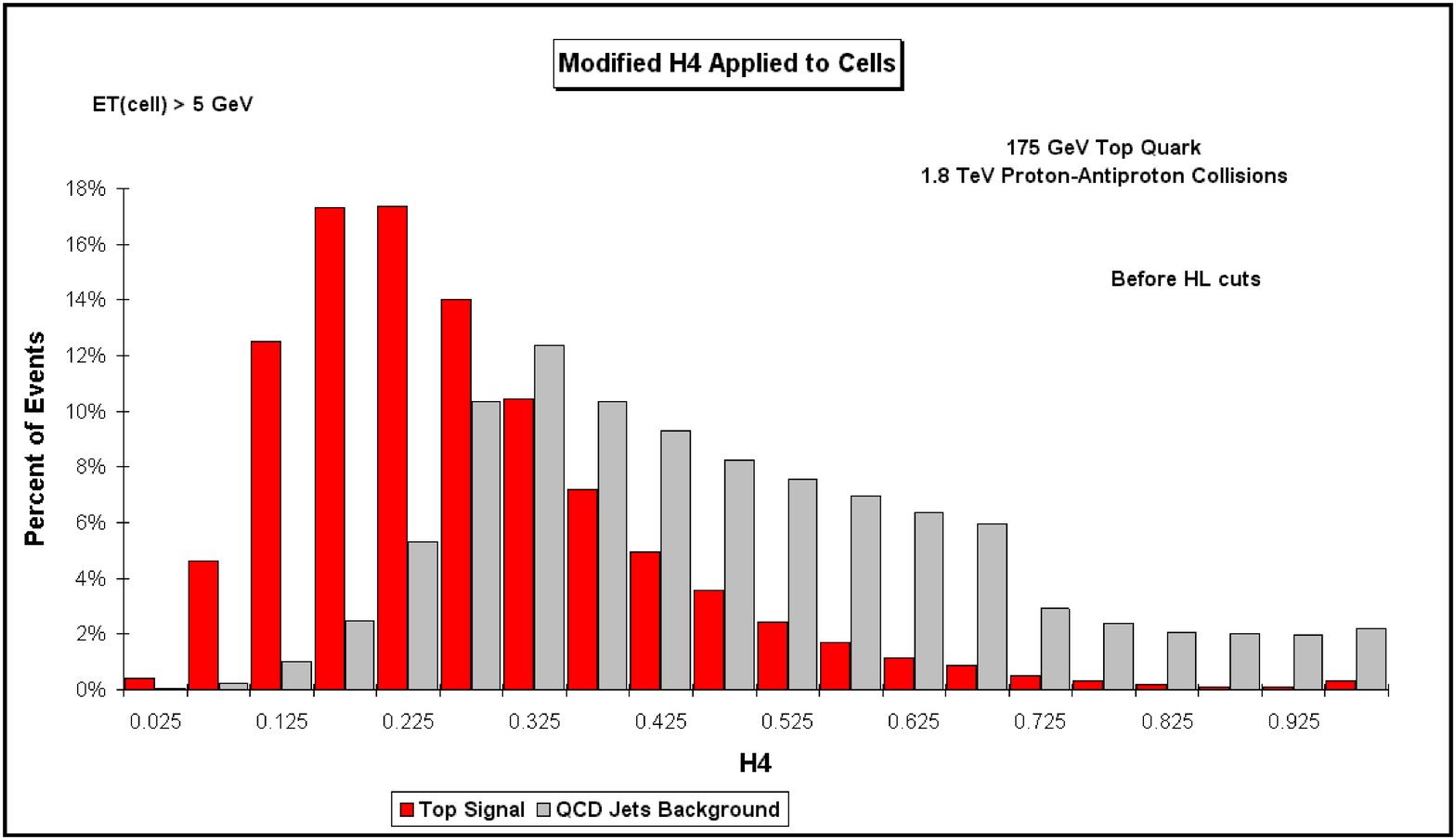}
  \end{center}
\caption{\footnotesize Shows the modified Fox-Wolfram moment, ${\hat H}_4$, calculated directly from the calorimeter cells with $E_T(cell)\!>\!5\gev$ for top-pair signal and for the QCD multi-jet background.  The plot shows the percentage of events in a $0.05$ bin with the sum of all bins normalized to $100\%$.}
\label{Fig:H4}
\end{figure}

\begin{figure}[htbp]
  \begin{center}
    \leavevmode
    \epsfxsize=4in 
    \epsfbox{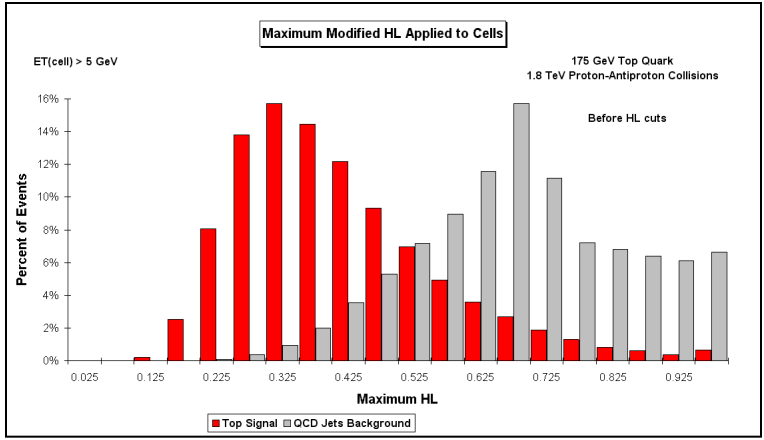}
  \end{center}
\caption{\footnotesize Shows the largest of the first six modified Fox-Wolfram moments, ${\hat H}_\ell$, for $\ell=1,\ldots,6$ in each event calculated directly from the calorimeter cells with $E_T(cell)\!>\!5\gev$ for top-pair signal and for the QCD multi-jet background.  The plot shows the percentage of events in a $0.05$ bin with the sum of all bins normalized to $100\%$.}
\label{Fig:Hmax}
\end{figure}

Table~\ref{Table:mean} shows the mean values and standard deviations for six of the modified Fox-Wolfram moments calculated using all cells with $E_T(cell)\!>\!5\gev$ for the top-pair signal and the QCD multi-jet background.   The mean values of the six moments ${\hat H}_1,\ldots ,{\hat H}_6$ are considerably smaller for the signal than the background.  For our calorimeter
( $3849$ cells with $\Delta\eta\Delta\phi=0.1\times7.5^\circ$) equal transverse energy in every cells yields ${\hat H}_\ell=0$ for odd $\ell$ and ${\hat H}_2=0.39$, ${\hat H}_4=0.23$, ${\hat H}_6=0.15$.  This corresponds to a cylindrically symmetric 
 ``blob".  The signal lies closer to this ``blob" configuration in ${\hat H}_\ell$-space than does most of the background events.  The background contains many two, three, and four jet configurations in addition to some higher jet multiplicity configurations.   The top-pair transverse energy deposition is usually more spread out in $\eta$-$\phi$ space than the background. This can be seen in Figs.~\ref{Fig:H2} , and \ref{Fig:H4} which show the ${\hat H}_2$,  and ${\hat H}_4$  distributions, respectively, for the signal and background.  In a given event, all six moments are, on the average, small for the top-pair signal, whereas for the background usually at least one of the moments is large.   This can be seen in Fig.~\ref{Fig:Hmax} which shows the distribution of the maximum of the six moments, ${\hat H}_1,\ldots,{\hat H}_6$, in each event for the top-pair signal and the QCD multi-jet background.  

\subsection{Constructing Modified Fox-Wolfram Moments from Jets}
Instead of using the calorimeter cells directly to characterize the event topology one can define ``jets" and use them to construct modified Fox-Wolfram moments.  We define jets using a simple algorithm.  One first considers the ``hot'' cells (those with transverse energy greater than $5\gev$).  Cells are combined to form a jet if they lie within a specified ``radius''  $R_j^2={\Delta\eta}^2+{\Delta\phi}^2$ in $\eta$-$\phi$ space from each other. Jets have an energy given by the sum of the energy of each cell in the cluster and a momentum $\vec p_j$ given by the vector sum of the momentums of each cell.  The invariant mass of a jet is simply $M_j^2=E_j^2-{\vec p}_j\cdot{\vec p}_j$.  In this analysis, we examine both ``narrow", $R_j=0.4$, and ``fat", $R_j=0.7$, jets, where jets are required to have at least $15\gev$ of transverse energy. 

The modified Fox-Wolfram moments are constructed from jets as follows,
\beq
{\hat H}_\ell(jets)=\left({4\pi\over 2\ell+1}\right)
\sum^{+\ell}_{m=-\ell}
\Bigl|\sum_i^{jets} Y^m_\ell(\Omega_i){E_T^i\over E_T({\rm sum})}\Bigr|^2,\eeq
where the inner sum is now over all the jets in the event with transverse energy, $E_T^i$, greater than some minimum (which we take to be $15\gev$) and $\Omega_i=(\theta_i,\phi_i)$ are the angular locations of the jets. Here, $E_T(sum)$ is the sum of the transverse energy of all the jets that are included in the sum. 

Table~\ref{Table:mean} shows the mean values and standard deviations for six of the modified Fox-Wolfram moments calculated using all jets with $R_j=0.4$ and $E_T(jet)\!>\!15\gev$ for the top-pair signal and the QCD multi-jet background.   The mean values are similar to those constructed directly from the cells and as before the mean values of the six moments ${\hat H}_1,\ldots ,{\hat H}_6$ are considerably smaller for the signal than for the background.

One can use the modified Fox-Wolfram moments constructed either from the cells or from jets.  In either case the ${\hat H}_\ell$'s characterize the topology of the event. At this point one could make a simple cut on ${\hat H}_\ell(max)$ to enhance signal over background (see Fig.~\ref{Fig:Hmax}), but one can do better by considering all six moments.  The six  moments ${\hat H}_1,\ldots ,{\hat H}_6$ form a six dimensional space in which different regions of the space correspond to different event topologies.  They range from zero to one and make excellent inputs into a neural network or Fisher discriminate \cite{rdf_top}.   In this paper, we will restrict events to lie within a region of the six dimensional ${\hat H}_\ell$-space.  The region will be defined by $L_\ell < {\hat H}_\ell < R_\ell$ for $\ell=1,\ldots,6$.  The left, $L_\ell$, and right, $R_\ell$, cuts will be selected using a genetic algorithm (GA) to maximize the signal over the square root of the background.  

\vskip 0.2in
\section{Multi-Dimensional Linear Cuts and Genetic Algorithms}

Genetic Algorithms are a broad class of minimization algorithms modeled after genetics and evolution \cite{Holland, Goldberg}.  In this paper, we will use a GA to perform ``optimal'' multi-dimensional linear cuts. In particular, we are interested in finding a set of left, $L_\ell$, and right, $R_\ell$, cuts ($\ell=1,\ldots,6$) that maximizes the signal, $N_{sig}$, over the square root of the background, $\sqrt{N_{bak}}$ (\ie the statistical significance). 

Unlike local algorithms, such as the Gradient Descent algorithm, GA's are much less likely to find and stay in a local minimum.  This is a considerable advantage for a large class of problems, including our particular application.  At the same time, GA's have local properties which make it possible to find and refine the ``optimal'' solutions in a reasonable time, while at the same time not precluding the possibility that there might be an even better solution.

\begin{figure}[htbp]
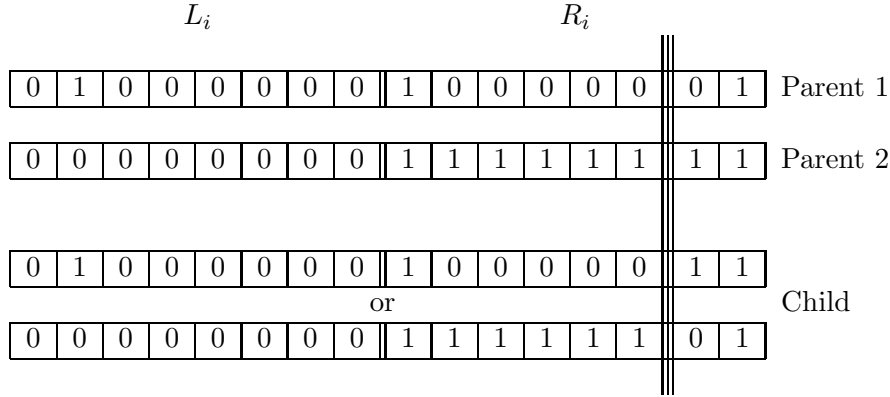

\begin{center}
\begin{tabular}{|c|c|c|c|c|c|c|c||c|c|c|c|c|c|||c|c|l}  
\multicolumn{8}{c}{$L_i$} & \multicolumn{8}{c}{$R_i$} 		& 		\\
\multicolumn{14}{c|||}{}  & \multicolumn{2}{c}{}      		& 		\\ \cline{1-16}
0 & 1 & 0 & 0 & 0 & 0 & 0 & 0 & 1 & 0 & 0 & 0 & 0 & 0 & 0 & 1 	& Parent 1	\\ \cline{1-16}
\multicolumn{14}{c|||}{} & \multicolumn{2}{c}{} 		& 		\\ \cline{1-16}
0 & 0 & 0 & 0 & 0 & 0 & 0 & 0 & 1 & 1 & 1 & 1 & 1 & 1 & 1 & 1 	& Parent 2	\\ \cline{1-16}
\multicolumn{14}{c|||}{} & \multicolumn{2}{c}{} 		& 		\\
\multicolumn{14}{c|||}{} & \multicolumn{2}{c}{} 		& 		\\ \cline{1-16}
0 & 1 & 0 & 0 & 0 & 0 & 0 & 0 & 1 & 0 & 0 & 0 & 0 & 0 & 1 & 1 	& 		\\ \cline{1-16}
\multicolumn{2}{c}{} & \multicolumn{12}{c|||}{or} & \multicolumn{2}{c}{} & Child\\ \cline{1-16}
0 & 0 & 0 & 0 & 0 & 0 & 0 & 0 & 1 & 1 & 1 & 1 & 1 & 1 & 0 & 1 	& 		\\ \cline{1-16}
\multicolumn{14}{c|||}{} & \multicolumn{2}{c}{} 		& 		\\
\end{tabular}
\end{center}
\caption{\footnotesize Crossover of two parental genes.  A split position is chosen at random within the genes of the parents.  The child receives all the bits to the left from one parent and all the bits to the right from the other parent. Shows the two genes $L_i\!=\!0.251$ and $R_i\!=\!0.5059$ for one parent and $L_i\!=\!0.0$ and $R_i\!=\!1.0$ for the other parent .  For this crossover, the children receive an unchanged $L_i$ (one gets $L_i\!=\!0.251$ and the other gets  $L_i\!=\!0.0$) and both get a modified $R_i$ that is a combination of the parental bits ($R_i\!=\!0.5137$ for the one child and $R_i\!=\!0.9922$ for the other).}
\label{Fig:crossover}
\end{figure}

To use a GA, one must have a set of data and a parametric real valued function on that data.  In our case, the data is the set of  six modified Fox-Wolfram moments ${\hat H}_1,\ldots ,{\hat H}_6$ for $10,000$ top-pair signal events and $10,000$ QCD multi-jet background events. The real valued function, $R_f$, on the data is the number of signal events over the square root of the number of background events that lie within a region of the six dimensional ${\hat H}_\ell$-space defined by $L_\ell < {\hat H}_\ell < R_\ell$ for $\ell=1,\ldots,6$.  Namely,
\beq
R_f={N_{sig}\over\sqrt{N_{bak}}},{\rm where}\ L_\ell < {\hat H}_\ell < R_\ell, \ \ell=1,\ldots,6.\eeq

In biological terms, the set of signal and background events is the environment in which a population resides, the real valued function, $R_f$, is analogous to the overall fitness of an individual for survival and reproduction, and the $12$ parameters $L_\ell$ and $R_\ell$ ($\ell=1,\ldots,6$) are the genes of an individual.  Since each of the left, $L_\ell$, and right, $R_\ell$, cuts lie between zero and one, we can multiply them by $255$ and represent them as a single byte (eight bits) within the computer\footnote{In our calculations we use two bytes to represent each real number, but for illustration it is simpler to consider just one byte.}.  For example, the gene corresponding to the left cut $L_i=0.251$ is represented in the computer as follows:
\beq
L_i=0.251\ \rightarrow\ [01000000].\eeq
Each individual has a set of $12$ genes corresponding to the $6$ pairs of left and right cuts. These $12$ genes form the ``DNA" of the individual which is represented in the computer as a string of  $12$ bytes.  For example, all left cuts of zero and all right cuts of one looks like the following:
\beq
L_1,R_1,\ldots,L_6,R_6\ \rightarrow\ [00000000][11111111]\dots[00000000][11111111].\eeq

Finding  an ``optimal'' solution or solutions is achieved through genetic evolution of a population over many generations. Typically, we use $500$ to $1,000$ individuals with an average life span of several simulation years and evolve them through $50$ to $100$ generations.  The genetic evolution of the population is achieved through the following mechanisms:

\begin{itemize}
 \item{\bf Natural selection:} At the end of each simulation year, the individuals with the worst performance are given the highest chance of dying and, therefore, their effect on future generations is minimized.  We do not, however, exterminate the worst performers unconditionally as is sometime done.  This usually decreases the convergence property of the GA, since ``good'' genes often require time before they lead to ``optimal'' results.

 \item{\bf Reproduction:} Each simulation year, depending on the population size, individuals reproduce by selecting a mate.  Individuals with higher performance have a higher probability of being selected, which further enhances the convergence property of the GA.  If the population is bigger, the rate of reproduction is smaller and vice versa.  This has the effect of better convergence because a population is small if many of the individuals do not perform well (which happens either at the initial stage of training, or when an already trained population discovers a new, much better solution, which makes all other individuals bad performers). During reproduction, the following two factors are critical:

\begin{itemize}
 \item{\bf Crossover:} The new individual inherits certain genes from one parent and others from the other.  This has the effect on both global and local property of the GA, since on one hand the ``good'' genes are preserved (local), while on the other hand new combinations are formed which has the effect of spanning the entire parameter space (global).  The probability of a crossover is determined by the crossover rate, $R_c$.

 \item{\bf Mutation:} The new individual has some of its genes randomly modified.  This is an extremely important factor in GA's, since this is the primary mechanism of discovering radically new solutions, which, if good, eventually start dominating the population.  If not good, the individuals that carry them die earlier due to natural selection and also due to the smaller probability for reproduction. The probability of a mutation is determined by the mutation rate, $R_m$.  The lower the mutation rate, the more local the GA is and vice versa.  

\end{itemize} 
\end{itemize} 

In the computer both {\em crossover} and {\em mutation} are bit-level operations on the genes (bytes). For example, Fig.~\ref{Fig:crossover} shows the crossover of genes from two parents. A split position is selected at random and the child inherits all bits before the split from one of the parents, and all bits after the split from
the other. The gene affected by the split becomes a combination the parental genes, while all other genes are inherited unaltered. This mechanism has the effect of both local convergence to an already ``good'' solution (if the high bits of both parents are the same), and at the same time exploring new combinations on {\em one} of the parameters only.  After {\em crossover}, the child's genes are mutated by randomly changing some of its bits. The probability for mutation is usually small in order to allow natural selection and crossover to find the best gene combinations in the already existing genetic pool of the population.  However, the mutation rate should not be zero in order to continuously probe the entire space of parameters for potentially better solutions that are not yet part of the genetic pool.

For most GA implementations the crossover rate, $R_c$, and the mutation rate, $R_m$, are {\em fixed} parameters.  However, this can result in poor convergence properties, since these rates (especially the mutation rate), have different effects on a non-trained population and an already trained one.  During the first generations of training, it is desirable that these rates be high in order to quickly scan the parameter space {\em globally}.  At later stages, when the genetic pool of the population is ``good'', too high rates interfere with the preservation of the genetic pool and the performance is detrimented significantly.  A ``fine-tuning'' of these rates as a function of simulation time is impractical, so instead we let both the crossover rate and the mutation rate be genes themselves.  In other words, their value at any point in time is subject to the same evolution as the parameters of the problem itself. This has a dramatic positive effect on the convergence property of the GA.  Initially, when the genetic pool is random, the values of the crossover and mutation rates are very high ($50\%$ on the average).  This allows for a very fast global scanning of the entire parameter space.  As the genetic pool improves, individuals with genes corresponding to high crossover and mutation rates (even if they are good performers otherwise), produce offspring which significantly deviates from the parental genes and, in all likelihood, does not perform as well.  In subsequent simulation years, this offspring is disadvantaged due to natural selection as well as mate selection and thus its genes are not likely to be passed on to future generations. On the other hand, individuals with good performance {\em and} reasonable crossover and mutation rates are more likely to produce offspring reflecting their genetic make-up and, therefore, have a much higher probability for their offspring surviving and reproducing.  This process is dynamic and the population constantly changes the crossover and mutation rates.

Furthermore, after {\em crossover} and {\em mutation}, we shift {\em one} of the left, $L_\ell$, or right, $R_\ell$, cuts by an amount $\Delta$,
\beq
L_\ell\to L_\ell\pm\Delta\ {\rm or}\  R_\ell\to R_\ell\pm\Delta.\eeq
This constitutes an implicit local algorithm, which allows for faster refinement of the solution and it further enhances the convergence properties of the GA algorithm.  In addition, we let the value of $\Delta$ also be a gene so that the complete ``DNA" of an individual consists of  the $15$ genes, $R_m$, $R_c$, $\Delta$, $L_1$, $R_1$,\ldots, $L_6$, $R_6$.  The population discovers dynamically the best values for the mutation rate, $R_m$, the crossover rate, $R_c$, and $\Delta$.  In particular, during the initial stages of training, $\Delta$ is totally irrelevant, since the major force of change is mutation. During later stages, the implicit local algorithm helps to refine an already good solution. At the final stages of evolution, the crossover and mutation rates are very low, and the improvement is dominated by $\Delta$, until eventually even the local algorithm cannot improve the solution any more.  In that case, $\Delta$ itself becomes very small.  This constitutes the criterion that further evolution is not likely to improve the population any further.  

Table~\ref{Table:cuts} shows the left, $L_\ell$, and right, $R_\ell$, cuts ($\ell=1,\ldots,6$) on the ${\hat H}_\ell$'s determined from our genetic algorithm (GA) procedure to maximize signal over the square root of the background.  We consider three cases.  In the first case the modified Fox-Wolfram moments are constructed directly from the calorimeter cells, ${\hat H}_\ell(cell)$, with $E_T(cell)>5\gev$.  The other two cases are for modified Fox-Wolfram moments constructed  from ``narrow", $R_j\!=\!0.4$, jets and from ``fat", $R_j\!=\!0.7$, jets, ${\hat H}_\ell(jet)$, with $E_T(jet)>15\gev$.

\begin{table}[htbp]
\begin{center}
\begin{tabular}{||l||c|c|c||} \hline \hline
 & ${\hat H}_\ell(cell)$ Cuts & ${\hat H}_\ell(jet)$ Cuts $R_j\!=\!0.4$ & ${\hat H}_\ell(jet)$ Cuts $R_j\!=\!0.7$ \\ 
 & \ Left (L) \ \ Right (R) & \ Left (L) \ \ Right (R) & \ Left (L) \ \ Right (R) \\ \hline\hline
 ${\hat H}_1$ & 0.000198 \ \ 0.347951 & 0.000000 \ \ 0.216602 & 0.007355 \ \ 0.217731 \\ \hline
 ${\hat H}_2$ & 0.011261 \ \ 0.225223 & 0.000000 \ \ 0.218647 & 0.000000 \ \ 0.256138 \\ \hline
 ${\hat H}_3$ & 0.013932 \ \ 0.249973 & 0.000000 \ \ 0.265553 & 0.009720 \ \ 0.160235 \\ \hline
 ${\hat H}_4$ & 0.000565 \ \ 0.588556 & 0.043092 \ \ 0.381796 & 0.021011 \ \ 0.491890 \\ \hline
 ${\hat H}_5$ & 0.051927 \ \ 0.192233 & 0.000000 \ \ 0.288945 & 0.018845 \ \ 0.395163 \\ \hline
 ${\hat H}_6$ & 0.026032 \ \ 0.912840 & 0.081071 \ \ 0.726467 & 0.026642 \ \ 0.794415 \\ \hline\hline
\end{tabular}
\end{center}
\caption{\footnotesize Shows the ${\hat H}_\ell$ cuts determined from a genetic algorithm (GA) to maximize the signal over the square root of the background.  The ${\hat H}_\ell$'s are restricted to lie in the region  $L_\ell < {\hat H}_\ell < R_\ell$ for $\ell=1,\ldots,6$ and are constructed from the calorimeter cells directly, ${\hat H}_\ell(cell)$, or from ``narrow", $R_j\!=\!0.4$, jets and from ``fat", $R_j\!=\!0.7$, jets(${\hat H}_\ell(jet)$).}
\label{Table:cuts}
\end{table}

\vskip 0.2in
\section{Reconstructing the Top-Pair Invariant Mass}

The top-pair invariant mass, $M_{t\tbar}$, corresponds to the center-of-mass energy, $\hat E_{cm}$, of the underlying parton-parton two-to-two subprocess which has a threshold at twice the mass of the top quark, $\hat E_{cm}\ge 2M_{top}$.  Although one cannot precisely reconstruct the parton-parton CM energy, the hope is that one will be able to observe a peak in the reconstructed top-pair invariant mass at twice the top quark mass. The size of this peak relative to the background determines whether this mode can be seen.  The top-pair-invariant mass can be reconstructed  from the outgoing jets or directly from the calorimeter cells.

\subsection{Using the Calorimeter Cells Directly}
The parton-parton invariant mass can be constructed directly form the calorimeter cells as follows:
\beq
M_{t\tbar}^2=E_{cells}^2-{\vec p}_{cells}^{\ 2} \ ,\eeq
where
\beq
{\vec p}_{cells}=\sum_i^{cells}{\vec p}_i,\eeq
and
\beq
E_{cells}=\sum_i^{cells} E_i.\eeq
The overall cell energy, $E_{cells}$, and momentum, ${\vec p}_{cells}$, is constructed by summing over all cells with transverse energy greater that some minimum (witch we take to be $5\gev$).

\subsection{Using the Outgoing Jets}
The top-pair invariant mass, $M_{t\tbar}$, can be constructed from the energy and momentum of the outgoing jets in the event as follows:
\beq
M_{t\tbar}^2=E_{jets}^2-{\vec p}_{jets}^{\ 2} \ ,\eeq
where
\beq
{\vec p}_{jets}=\sum_i^{jets}{\vec p}_i,\eeq
and
\beq
E_{jets}=\sum_i^{jets} E_i.\eeq
The overall jet energy, $E_{jets}$, and momentum, ${\vec p}_{jets}$, is constructed by summing over all jets with transverse energy greater than $15\gev$.

\begin{figure}[htbp]
  \begin{center}
    \leavevmode
    \epsfxsize=4in 
    \epsfbox{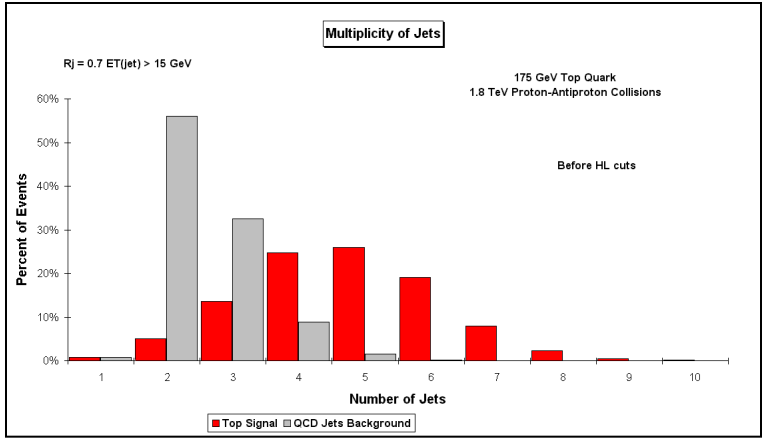}
  \end{center}
\caption{\footnotesize Shows the multiplicity of  ``fat" jets ($R_j\!=\!0.7$) with transverse energy greater than $15\gev$ for the top-pair signal and the QCD multi-jet background.  The plot shows the percentage of events with $N$ jets with $E_T(jet)>15\gev$.}
\label{Fig:Njet}
\end{figure}

\vskip 0.2in
\section{Isolating Multi-Jet Topologies}

\begin{table}[htbp]
\begin{center}
\begin{tabular}{||l||c|c|c|c|c|c||} \hline\hline
                  &{\footnotesize  Mass Type}            & {\footnotesize Mass Range}        
& {\footnotesize $N_{sig}$} 
&{\footnotesize  $N_{bak}$} & {\footnotesize $N_{bak}/N_{sig}$} & {\footnotesize $N_{sig}/\sqrt{N_{bak}}$} \\ \hline\hline
 Jet  Cuts $N_j\!\ge\!5$ & {\small Jet Mass}   & {\footnotesize $>\!300\gev$} & 364    & 444,551    
& 1,221   & 0.55 \\ 
 {\footnotesize $R_j\!=\!0.7$, $E_T(jet)\!>\!15\gev$} &         &  	       &        &            &         & \\ \hline
 ${\hat H}_\ell(cell)$ Cuts & {\small Cell  Mass}  & {\footnotesize $>\!250\gev$} & 54     & 4,621      & 85      & 0.80 \\ 
 {\footnotesize $E_T(cell)\! >\!5\gev$} &        &             &        &            &         & \\ \hline
 ${\hat H}_\ell(cell)$ Cuts & {\small  Jet  Mass}   & {\footnotesize $>\!300\gev$} & 65     & 8,138      & 125     & 0.72 \\ 
{\footnotesize $R_j\!=\!0.4$, $E_T(cell)\!>\!5\gev$}        &  &    &        &            &         & \\ \hline
 ${\hat H}_\ell(jet)$  Cuts & {\small  Jet Mass}   & {\footnotesize $>\!300\gev$} & 105    & 17,578     & 168     & 0..79 \\ 
{\footnotesize $R_j\!=\!0.4$, $E_T(jet)\!>\!15\gev$}      &  &     &        &            &         & \\ \hline
 ${\hat H}_\ell(jet)$  Cuts & {\small  Jet Mass}   & {\footnotesize $>\!300\gev$} & 87     & 31,843     & 365     & 0.49 \\ 
{\footnotesize $R_j\!=\!0.7$, $E_T(jet)\!>\!15\gev$}        &  &        &        &            &         & \\ \hline\hline
\end{tabular}
\end{center}
\caption{\footnotesize $175\gev$ top quark pairs produced in $1.8\tev$ proton-antiproton collisions.  The table shows the number of events (with ${\cal L}=100$/pb) for the top-pair signal and the QCD multi-jet background remaining after making a jet multiplicity cut ($N_j\ge 5$, $R_j\!=\!0.7$, $E_T(jet)>15\gev$ and after making various ${\hat H}_\ell$ cuts.  The ${\hat H}_\ell$'s are constructed from the calorimeter cells directly, ${\hat H}_\ell(cell)$, or from narrow, $R_j\!=\!0.4$, jets and from fat, $R_j\!=\!0.7$, jets, ${\hat H}_\ell(jet)$.  The ${\hat H}_\ell$'s are restricted to lie in the region  $L_\ell < {\hat H}_\ell < R_\ell$ for $\ell=1,\ldots,6$, where the left, $L_\ell$, and right, $R_\ell$, cuts are selected using a genetic algorithm (GA) to maximize the signal over the square root of the background and are given in Table~\ref{Table:cuts}. The top-pair invariant mass is calculated either directly from the cells ({\em cell mass}) or from the jets ({\em jet mass}).}
\label{Table:results}
\end{table}

\subsection{Using Jet Multiplicity Cuts}

Fig.~\ref{Fig:Njet} shows the multiplicity of jets, $N_j$, (with $R_j\!=\!0.7$ and $E_T>15$ GeV) for the top-pair signal and the QCD multi-jet background.  One obvious way to enhance the top-pair signal over the background is to demand the events to have a minimum number of jets, $N_j(min)$ (usually taken to be five).  Table~\ref{Table:results} shows that after a jet multiplicity cut  there are about $360$ signal events and roughly $460,000$ background events (in $100$/pb) for the reconstructed mass range $M_{t\tbar}>300\gev$.  The background is about a factor of $1,200$ times larger than the signal.

\begin{figure}[htbp]
  \begin{center}
    \leavevmode
    \epsfxsize=4in 
    \epsfbox{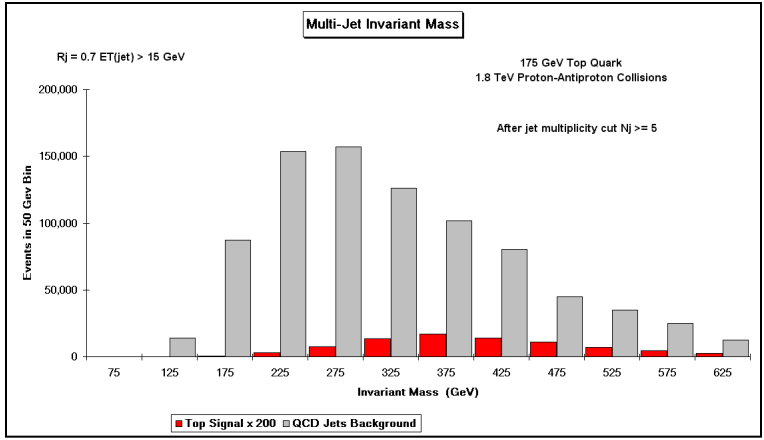}
  \end{center}
\caption{\footnotesize Shows the reconstructed top-pair invariant mass, $M_{t\tbar}$, for $175\gev$ top quarks produced in $1.8\tev$ proton-antiproton collisions together with the QCD multi-jet background for events that have survived the jet multiplicity cut, $N_j\ge 5$.  The invariant mass is constructed from all the jets ($R_j\!=\!0.7$) in the event with $E_T(jet)>15\gev$.  The plot shows the number of events (with ${\cal L}=100$/pb) in a $50\gev$ bin. The top-pair signal has been multiplied by a factor of $200$.}
\label{Fig:MjJcut}
\end{figure}

Fig.~\ref{Fig:MjJcut} shows the top-pair invariant mass reconstructed from the ``fat" jets in the event (with $E_T(jet)>15$ GeV) for the top-pair signal ({\em multiplied by $200$}) and the QCD multi-jet background after a jet multiplicity cut ($N_j\ge5$).  A problem that arises when using a jet multiplicity cut is that the cut causes an artificial peak in the background invariant mass near the peak in the signal.  Requiring a minimum number of jets with transverse energy greater than $15$ GeV removes events with low parton-parton invariant mass.  In addition, jet multiplicity cuts are ``quantized" (\ie discrete).  One cannot smoothly vary the degree of the cut to, for example, optimize signal over background.

\subsection{Using ${\hat H}_\ell$ Cuts Without Jets}

In this section, we will examine a method for isolating the top-pair signal over the background without defining jets at all.  The calorimeter cell information is used directly to select the events and to reconstruct the top-pair invariant mass. The six modified Fox-Wolfram moments
${\hat H}_1,\ldots ,{\hat H}_6$ constructed from the calorimeter cells, ${\hat H}_\ell(cells)$, are used to select events.  Events are required to lie in a region of ${\hat H}_\ell$-space defined by $L_\ell < {\hat H}_\ell < R_\ell$ for $\ell=1,\ldots,6$.  The left, $L_\ell$, and right, $R_\ell$, cuts given in Table~\ref{Table:cuts} were determined from our genetic algorithm procedure which maximize the signal over the square root of the background.  No jet multiplicity cuts are made.

\begin{figure}[htbp]
  \begin{center}
    \leavevmode
    \epsfxsize=4in 
    \epsfbox{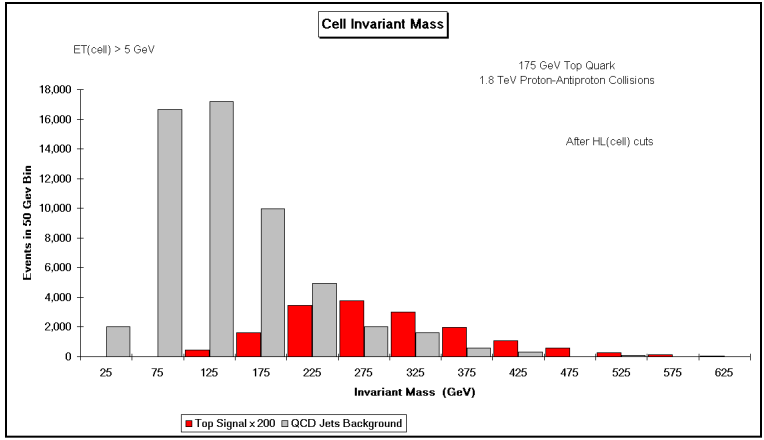}
  \end{center}
\caption{\footnotesize Shows the reconstructed top-pair invariant mass, $M_{t\tbar}$, for $175\gev$ top quarks produced in $1.8\tev$ proton-antiproton collisions together with the QCD multi-jet background for events that have survived the ${\hat H}_\ell(cell)$ cuts.  Events are required to have ${\hat H}_\ell$'s in the region  $L_\ell < {\hat H}_\ell < R_\ell$ for $\ell=1,\ldots,6$, where the left, $L_\ell$, and right, $R_\ell$ cuts are given in Table~\ref{Table:cuts}.  The ${\hat H}_\ell(cell)$'s and the invariant mass are constructed directly from the calorimeter cells using all cells in the event with $E_T(cell)>5\gev$. Jets are never defined and no jet multiplicity cuts are made.  The plot shows the number of events (with ${\cal L}=100$/pb) in a $50\gev$ bin. The top-pair signal has been multiplied by a factor of $200$.}
\label{Fig:McHcCut}
\end{figure}

Fig.~\ref{Fig:McHcCut} shows the top-pair invariant mass reconstructed directly from the calorimeter cells (with $E_T(cell)>5$ GeV) for the top-pair signal ({\em multiplied by 200}) and the QCD multi-jet background after the ${\hat H}_\ell$ cuts.  Table~\ref{Table:results} shows that for the reconstructed mass range $M_{t\tbar}>250\gev$ there are about $50$ signal events and roughly $5,000$ background events (in $100$/pb).  Here the background is about a factor of $100$ larger than the signal. 
For the top-pair signal, the invariant mass reconstructed from cells with $E_T(cell)>5\gev$ peaks at about $275\gev$ which is less than the {\em true} top-pair mass of $350\gev$.  Removing cells with transverse energy less than $5\gev$ reduces the reconstructed mass from its generated value.  Nevertheless, this method gives our best statistical significance of $0.8$.  One is looking for a bump above a smoothly falling background and it does not matter if the mass is shifted downward.  One can always correct the mass after one establishes the signal.

Furthermore, the use of this method, in principle, does not cause an artificial peak in the reconstructed invariant mass for the background.   There is a peak in the background in Fig.~\ref{Fig:McHcCut} but it is at much lower mass than the signal and could be eliminated altogether by lowering the minimum cell transverse energy of $5\gev$.

\begin{figure}[htbp]
  \begin{center}
    \leavevmode
    \epsfxsize=4in 
    \epsfbox{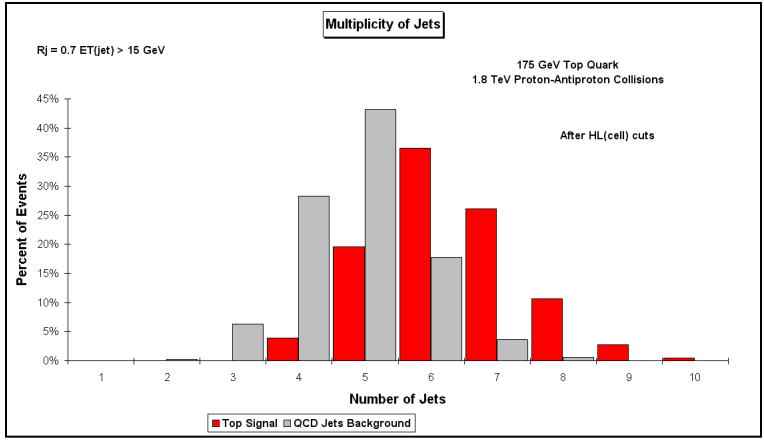}
  \end{center}
\caption{\footnotesize Shows the multiplicity of  ``fat" jets ($R_j\!=\!0.7$) with transverse energy greater than $15\gev$ for the top-pair signal and the QCD multi-jet background  for events that have survived the ${\hat H}_\ell(cell)$ cuts.  Events are required to have ${\hat H}_\ell(cell)$'s in the region  $L_\ell < {\hat H}_\ell(cell) < R_\ell$ for $\ell=1,\ldots,6$, where the left, $L_\ell$, and right, $R_\ell$ cuts are given in Table~\ref{Table:cuts}.  The plot shows the percentage of events with $N$ jets with $E_T(jet)>15\gev$.}
\label{Fig:NjHcCut}
\end{figure}

\begin{figure}[hp]
  \begin{center}
    \leavevmode
    \epsfxsize=4in 
    \epsfbox{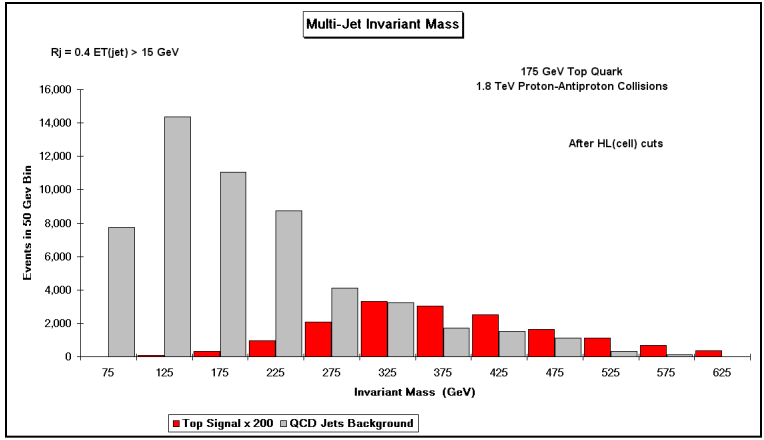}
  \end{center}
\caption{\footnotesize Shows the reconstructed top-pair invariant mass, $M_{t\tbar}$, for $175\gev$ top quarks produced in $1.8\tev$ proton-antiproton collisions together with the QCD multi-jet background for events that have survived the ${\hat H}_\ell(cell)$ cuts.  Events are required to have ${\hat H}_\ell(cell)$'s in the region  $L_\ell < {\hat H}_\ell < R_\ell$ for $\ell=1,\ldots,6$, where the left, $L_\ell$, and right, $R_\ell$ cuts are given in Table~\ref{Table:cuts}. The ${\hat H}_\ell(cell)$'s are constructed from all the cells in the event with $E_T(cell)>5\gev$ and the invariant mass is constructed from all the jets ($R_j\!=\!0.4$) in the event with $E_T(jet)>15\gev$.  No jet multiplicity cuts are made.  The plot shows the number of events (with ${\cal L}=100$/pb) in a $50\gev$ bin. The top-pair signal has been multiplied by a factor of $200$.}
\label{Fig:MjHcCut}
\end{figure}

After the events have been selected using the ${\hat H}_\ell(cell)$'s, one can construct and examine the jets in the event.  Fig.~\ref{Fig:NjHcCut} shows the multiplicity of  ``fat" jets ($R_j\!=\!0.7$) with transverse energy greater than $15\gev$ for the top-pair signal and the QCD multi-jet background  for events that have survived the ${\hat H}_\ell(cell)$ cuts.  By selecting events that lie in the region of ${\hat H}_\ell$ space given in Table~\ref{Table:cuts}, we have selected events with a large number of jets, but in a smooth way.  The background now peaks at five jets instead of the two jet peak in Fig.~\ref{Fig:Njet} and the signal and background jet multiplicities now look similar.  Fig.~\ref{Fig:MjHcCut} shows the top-pair invariant mass, $M_{t\tbar}$, reconstructed from jets for the top-pair signal and the QCD multi-jet background for events that have survived the ${\hat H}_\ell(cell)$ cuts.  The invariant mass is constructed from all the ``narrow" jets ($R_j\!=\!0.4$) in the event with $E_T(jet)>15\gev$.  No jet multiplicity cuts are made.  Here the invariant mass of the signal peaks at around $325\gev$ and Table~\ref{Table:results} shows that the statistical significance is only slightly lower than the cell invariant mass case.

\subsection{Using ${\hat H}_\ell$ Cuts With Jets}

Instead of working with the cells directly, one can define jets from the very beginning and do the whole analysis with the jets.  The six moments modified Fox-Wolfram moments
${\hat H}_1,\ldots ,{\hat H}_6$ constructed from the jets, ${\hat H}_\ell(jet)$, are used to select events.  Events are required to lie in a region of ${\hat H}_\ell$-space defined by $L_\ell < {\hat H}_\ell < R_\ell$ for $\ell=1,\ldots,6$.  Table~\ref{Table:cuts} gives the left, $L_\ell$, and right, $R_\ell$, cuts determined from the genetic algorithm (GA) procedure to maximize the signal over the square root of the background.   The results for both  ``narrow" jets ($R_j\!=\!0.4$) and  ``fat" ($R_j\!=\!0.7$) jet is given in Table~\ref{Table:results}.  The ``narrow" jets produce better results than the ``fat" jets.

\begin{figure}[htbp]
  \begin{center}
    \leavevmode
    \epsfxsize=4in 
    \epsfbox{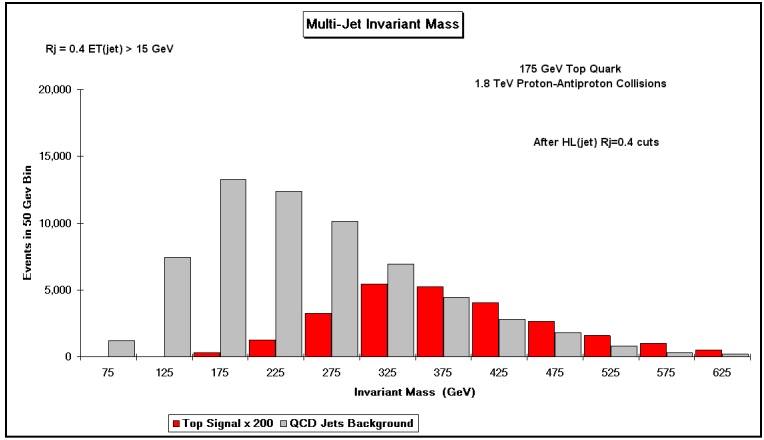}
  \end{center}
\caption{\footnotesize Shows the reconstructed top-pair invariant mass, $M_{t\tbar}$, for $175\gev$ top quarks produced in $1.8\tev$ proton-antiproton collisions together with the QCD multi-jet background for events that have survived the ${\hat H}_\ell(jet)$ cuts.  Events are required to have ${\hat H}_\ell(jet)$'s in the region  $L_\ell < {\hat H}_\ell(jet) < R_\ell$ for $\ell=1,\ldots,6$, where the left, $L_\ell$, and right, $R_\ell$ cuts are given in Table~\ref{Table:cuts}. The ${\hat H}_\ell(jet)$'s and the invariant mass are constructed from all the jets ($R_j\!=\!0.4$) in the event with $E_T(jet)>15\gev$, but no jet multiplicity cuts are made.  The plot shows the number of events (with ${\cal L}=100$/pb) in a $50\gev$ bin. The top-pair signal has been multiplied by a factor of $200$.}
\label{Fig:MjHjCut}
\end{figure}

Fig.~\ref{Fig:MjHjCut} shows the top-pair invariant mass reconstructed from the jets (with $E_T(jet)>15$ GeV and $R_j\!=\!0.4$) for the top-pair signal ({\em multiplied by 200}) and the QCD multi-jet background after the ${\hat H}_\ell(jet)$ cuts.  Table~\ref{Table:results} shows that for the reconstructed mass range $M_{t\tbar}>300\gev$ there are about $100$ signal events and roughly $17,000$ background events (in $100$/pb).  The background is about a factor of $170$ larger than the signal which is comparable to, but slightly worse than we get from using the cells directly.

\vskip 0.2in
\section{Summary and Conclusions}

It is difficult to completely isolate the six-jet decay mode of top-pair production over the QCD multi-jet background at hadron colliders without b-quark tagging.  We are able to reduce the background over the signal to less than a factor of $100$ using purely topological methods and without the use of  b-quark tagging.  B-quark tagging would, of course, further enhance the signal to background ratio. Our technique can be summarized as follows:

\begin{itemize}
\item{} Construct six modified Fox-Wolfram Moments, ${\hat H}_1,\ldots ,{\hat H}_6$, directly from the calorimeter cells or from jets.
\item{} Select events that lie in a certain region of ${\hat H}_\ell$-space defined by $L_\ell < {\hat H}_\ell < R_\ell$ for $\ell=1,\ldots,6$.  
\item{} Determine the left, $L_\ell$, and right, $R_\ell$, cuts using a genetic algorithm (GA) procedure that maximizes the signal over the square root of the background.  
\item{} Construct the top-pair invariant mass, $M_{t\tbar}$, directly from the calorimeter cells or from jets.
\end{itemize}

\noindent We do not make a jet multiplicity cut.  Jet multiplicity cuts cause an artificial peaking of the background invariant mass near the $2M_{top}$ peak of the signal, whereas requiring events to lie in a region of six-dimensional ${\hat H}_\ell$-space, in principle, does not.  Requiring the ${\hat H}_\ell$'s to be small does select events with a large number of jets, but in a smooth way.  Also, ${\hat H}_\ell$ cuts can be continuously varied, where jet multiplicity cuts are discrete.  Furthermore, the modified Fox-Wolfram moments, ${\hat H}_1,\ldots ,{\hat H}_6$, can be constructed directly from the calorimeter cells without the need to define jets.

We have used the six-jet decay mode of top-quark pair production hadron colliders as an example of our techniques.  Other parton-parton subprocesses can be isolated by selecting the regions of  ${\hat H}_\ell$-space that correspond to their unique topology.  For example, many super-symmetric subprocesses have characteristic event topologies where our techniques should also help to improve the signal to background ratio.

% numerical arg to 'thebibliography' is max width of an entry in characters

\end{document}